\nonstopmode
\documentclass[submission,copyright,creativecommons]{eptcs}
\providecommand{\event}{LFMTP'11} 
\usepackage{hyperref}
\usepackage{proof}
\usepackage[all]{xy}

\usepackage{xspace}
\usepackage{ifthen}
\usepackage{amsmath}
\usepackage{amsthm}
\usepackage{amssymb}
\usepackage{stmaryrd}

\usepackage{proof}

\newcommand{\abbrev}[1]{#1} 

\newcommand{\ie}{\abbrev{i.\,e.}}

\newcommand{\etal}{\abbrev{et.\,al.}}

\newcommand{\oann}[1]{{}^{#1}\kern-0.15ex}
\newcommand{\ovar}{\mathord{\bullet}}

\newcommand{\comment}[1]{} 
\newcommand{\sspace}{\,}
\newcommand{\lspace}{\ \,}
\newcommand{\shspace}{\vspace{1ex}}

\newcommand{\la}{\lambda}
\newcommand{\emptyVec}{[]}
\newcommand{\emptySet}{\emptyset}
\newcommand{\ve}[1]{[#1]}

\newcommand{\length}[1]{\mbox{length}(#1)}

\newcommand{\ApA}[2]{#1 \sspace #2}
\newcommand{\LaA}[2]{\la {#1}. \sspace #2}

\newcommand{\ApO}[3]{#1 \sspace ^{#2} \sspace #3}
\newcommand{\oapp}[1]{\sspace ^{#1} \sspace}
\newcommand{\LaO}[2]{\la ^{#1} . \sspace #2}

\newcommand{\freevars}[1]{\mathsf{fV}(#1)}

\newcommand{\multiinsert}[3]{#1 ^ {\sspace #3 / #2}}

\newcommand{\Val}[3]{ (\la^{#1} . \sspace {#2} )^{#3}}

\newcommand{\pa}[1]{\left( #1 \right)}
\newcommand{\sub}[2]{\left[ #1 / #2 \right]}
\newcommand{\subs}[4]{\left[ #1 / #2 , #3 / #4 \right]}
\newcommand{\multisubs}[4]{\left[ #1 / #2 , \ldots, #3 / #4 \right]}

\newcommand{\re}{\longrightarrow}

\newcommand{\ev}[2]{\llbracket{#1}\rrbracket_{#2}}
\newcommand{\ap}{\,@\,}

\newcommand{\parseb}[4]{{#1 \lhd #2 \rhd  #3[#4]}}

\newcommand{\parseFun}[1]{{\left(\mathord{\lhd} {#1} \mathord{\rhd}\right)}}
\newcommand{\parseFunE}{\mathop{\rhd}}

\newcommand{\infnamed}[3]{{\infer[#1]{#3}{#2}}}

\newcommand{\x}{X} 
\newcommand{\te}{T} 
\newcommand{\ot}{oT} 
\newcommand{\otw}{oT_\Gamma} 
\newcommand{\otc}{\overline{oT}} 
\newcommand{\va}{V} 
\newcommand{\xl}{\vec \x} 
\newcommand{\vl}{\vec \va} 
\newcommand{\gl}{\vec \Gamma} 

\newcommand{\pr}[2]{{ \prE   \left( #1 \ , \  #2 \right)}}
\newcommand{\prE}{{\lhd \hspace{-6pt} \lhd}}
\newcommand{\prVal}[1]{{\lhd \hspace{-2pt} \left( #1 \right)}}
\newcommand{\prValE}{{\mathord{\lhd}}}

\newcommand{\lab}[1]{\mbox{(#1)}}

\newcommand{\vstart}{\vec v_{start}}
\newcommand{\vrest}{\vec v_{rest}}

\newcommand{\define}[1]{\mbox{\textbf{\textit{#1}}}}

\title{A Lambda Term Representation \\ Inspired by Linear Ordered Logic}
\author{Andreas Abel
\institute{
Theoretical Computer Science\\
Institut f\"ur Informatik\\
Ludwig-Maximilians-Universit\"at\\
M\"unchen, Germany}
\email{andreas.abel@ifi.lmu.de}
\and
Nicolai Kraus
\institute{
Functional Programming Laboratory\\
School of Computer Science\\
University of Nottingham\\
Nottingham, United Kingdom}
\email{ngk@cs.nott.ac.uk}
}
\def\titlerunning{A Lambda Term Representation Inspired by Linear Ordered Logic}
\def\authorrunning{Andreas Abel and Nicolai Kraus}
\begin{document}
\maketitle

\begin{abstract}
We introduce a new nameless representation of lambda terms inspired by
ordered logic.  At a lambda abstraction, number and relative
position of all occurrences of the bound variable are stored, and
application carries the additional information where to cut the
variable context into function and argument part.  This way, 
complete information about free variable occurrence is available at each
subterm without requiring a traversal, and environments can
be kept exact such that they only assign values to variables that
actually occur in the associated term.
Our approach avoids space leaks in interpreters that build 
function closures.  

In this article, we prove correctness of the new representation and
present an experimental evaluation of its performance in a proof
checker for the Edinburgh Logical Framework.
  

\end{abstract}

\section{Introduction}
\label{sec:intro}

Type checking dependent types in languages like Agda~\cite{norell:PhD}
and Coq~\cite{inria:coq83} or logical frameworks like Twelf~\cite{carsten:twelf}
requires a large amount of evaluation, since types may depend on
values.  Such type checkers incorporate
an interpreter for purely functional programs with free variables---at least, the
$\lambda$-calculus---which is used to compute weak head normal forms
of types.  Efficiency of type checking is mostly identical with
efficiency of evaluation\footnote{Evaluation is necessary to reduce
  types to weak head normal form and compare types for equality.
  Subtracting these operations, type checking has linear complexity.
} 
(and, in case of type reconstruction,
efficiency of unification), and remains a challenge as of today.  
In seminal work, Gregoire and Leroy \cite{gregoireLeroy:icfp02} have
sped up Coq type checking by compiling to byte-code instead of
using an interpreter.  Boespflug \cite{boespflug:padl10} has obtained
further speed-ups by producing native code using stock-compilers.  

While compilation approaches are successful on batch type \emph{checking}
fully explicit programs, they have not been attempted on type
\emph{reconstruction} using higher-order unification or on interactive
program construction such as in Agda and Epigram \cite{mcBrideMcKinna:view}.
These languages are involved and constantly evolving, 
and their implementations are
prototypes and frequently modified and extended.  Implementing a
full compiler just to get type reconstruction going is deterring;
furthermore, compilation has not (yet) proven its feasibility in minor
evaluation tasks (like weak head evaluation)
that dominate higher-order unification.  
At least for language prototyping, smart
interpreters are, and may remain, competitive with compilation.  

For instance, Twelf's interpreter is sufficiently fast; it is 
inspired by a term representation with de Bruijn indices 
\cite{deBruijn:nameless} and explicit substitutions 
\cite{abadiCardelliCurienLevy:jfp91}.  In the context of functional
programming, explicit substitutions are known as \emph{closures},
consisting of the code of a function plus an environment, assigning
values to the free variables appearing in the code.  
In typical implementations of interpreters \cite{coquand:type}, 
these environments are not
precise; they assign values to all variables that are statically in
scope rather than only to those that are actually referred to in the
code.  This bears potential for space-leaks: the environment of a
closure might refer to a large value that is never used, but cannot be
garbage collected.  An obvious remedy to this threat is, when forming
a closure, to restrict the environment to the actual free variables;
however, this requires a traversal of the code.  We explore a
different direction: we are looking for a code representation that
maintains information about the free variables at each node of the
abstract syntax tree.

A principal candidate is \emph{linear typing} in Curry-Howard
correspondence with Girard's linear logic \cite{girard:linear}; there,
each variable in scope is actually referenced (more precisely,
referenced exactly once).  In other words, the
free variables are exactly the variables in the typing context.
Dropping types, we may talk of \emph{linear scoping}.
Yet we do not want to represent linear terms, but arbitrary
$\lambda$-terms.   Kesner and Lengrand \cite{kesnerLengrand:infcomp07}
achieve this by introducing explicit term constructs for weakening and
contraction.  We pursue a different path: we incorporate information
about variable use and multiplicity directly into abstraction and
application. 

In the context of linear $\lambda$-calculus, the free
variables of a function application are the disjoint union of the free
variables of the function and the free variables of the argument.  If
we want to maintain the set of free variables during a term traversal,
at an application node we need to decide which variables go into the
function part and which into the argument part.  Thus, we would store
at each application a set of variables that go into the, say, function
part, all others would go to the argument part.  Less information is
needed if we switch to an \emph{ordered} representation.

\emph{Ordered logic}, also called \emph{non-commutative linear logic}
\cite{polakowPfenning:tlca99}, refines linear logic by removing the
structural rule \emph{exchange} which restricts hypotheses to be used
\emph{in the order they have been declared}.  Transferring this
principle to ordered scoping this means that the scoping context lists
the free variables in the order they occur in the term, from left to
right.  This allows pushing the context into an application with very
little information: we just need to know \emph{how many} variables
appear in the function part so we can cut the context in two at the
right position, splitting it into function context and argument
context.  This constitutes the central idea of our representation: at
each application node of the syntax tree, we store a number denoting
the number of free variable occurrences in the function part.  During
evaluation of an application in an environment, we can cut the
environment into two, the environment needed for the evaluation of the
function and the environment needed for the evaluation of the
argument.  Thus, our environments are precise and space leaks are
avoided.  In particular, a variable is always evaluated in a singleton
environment assigning only a value to this variable.  Following this
observation, variables do not need a name, they are identified by
their position; and environments are simply sequences of values.

Since we are not interested in proof terms of ordered logic per se,
but only borrow the \emph{ordered context} idea for our
representation of untyped $\lambda$-calculus, 
we need to allow multiple occurrences of the same
variable.  In fact, the context shall list the variable
\emph{occurrences} in order.  At a lambda abstraction, we bind all
occurrences of the same variable.  Thus, at an abstraction node we
specify at which positions the bound variable should be inserted in
the scoping context.  This concludes the presentation of our idea.  

In the rest of the paper, after an introductory example
(Section~\ref{sec:example}) we formally define our term representation
in Section~\ref{sec:syntax}.   Interpreter and handling of
environments are described in Section~\ref{sec:values}, followed by
the translation between ordinary lambda terms and ordered terms
(Section~\ref{sec:parsing}).  Soundness of the interpreter is formally
proven in Section~\ref{sec:sound}, before we conclude with an
experimental evaluation in Section~\ref{sec:experiments}.

This article summarizes the B.Sc.\ thesis of the second author
\cite{kraus:bachelor}.

\section{An Example}
\label{sec:example}



To demonstrate the discussed risk of space-leaks during evaluation, we apply the term $\LaA x {\LaA y {\ApA{\ApA a b} y}}$ in basic syntax consecutively to the free variables $g$ and $f$. A possible (and, if the mentioned closures are used, typical) sequence of reduction steps is given below.  
By writing $t\multisubs {s_1} {x_1} {s_n} {x_n}$, we want to express that in the term $t$, each occurrence of the variable $x_1$ ($x_2, \ldots, x_n$) has to be replaced by the term $s_1$ (resp. $s_2, \ldots, s_n$) simultaneously. Such a substitution list always applies only to the directly preceding term:
\[
\begin{array}{lr@{}l}
\shspace &(\LaA x {\LaA y {\ApA{\ApA a b} y}}) & \sspace g \sspace f \\ 
\shspace\re &(\LaA y {\ApA{\ApA a b} y}) & \sub g x \lspace f \\ 
\shspace\re &      (\ApA{\ApA a b} y) & \subs g x f y \\ 
\shspace\re &      \multicolumn 2 l {(a\sspace b)\subs g x f y \lspace y\subs g x f y} \\ 
\shspace\re &      \multicolumn 2 l {a\subs g x f y \lspace b\subs g x f y \lspace f} \\ 
\shspace\re &      \multicolumn 2 {c} {a \sspace b \sspace f}  \\ 
\end{array}
\]
Here, the substitution $\sub g x$ could be dropped instantly and there is no need to apply the other substitution $\sub f y$ to the term $\ApA a b$. 
However, the term representation used above comes along with the problem that such an evaluation algorithm does not have the required information in time.
This is due to the fact that the binding information is always split between the $\la$ and the actual variable occurrence, as they both carry the variable name.  
In contrast, using de Bruijn indices would make it possible to remove the piece of information from the $\la$.
Our goal is to do it the other way round: We want the whole information to be available at the $\la$, thus making it possible to know the number and places of the bound variable occurrences without looking at the whole term.



\section{Syntax}
\label{sec:syntax}

In this article, we only cover the core constructs of the lambda calculus as they are enough to make the approach clear. However, we do not see any limitations for common extensions. We first define \define{ordered preterms}:
\[
\begin{array}{lllrll@{\qquad}}
\mathsf{ordered \ preterms}       & \ni & t,u & ::= & x & \mbox{free variable (named $x$)} \\
			                   &&& \mid & \ovar & \mbox{bound variable (nameless)} \\
			                   &&& \mid & \ApO t m u & \mbox{application} \\
			                   &&& \mid & \LaO {\vec k} t & \mbox{abstraction} \\
\end{array}
\]
\emph{Free variables} are denoted by their name like in the standard syntax. \emph{Bound variables}, however, are just denoted by a dot $\ovar$, which does not carry any information beside the fact that it is a bound variable. 
In the case of an \emph{application}, there is a first term (the function part) and a second term (the argument) as usual. 
Furthermore, the application carries an integer $m$ as an additional piece of information that will be important for the evaluation process and is explained 
in a moment.
The most interesting part is the \emph{abstraction} $\LaO {\vec k} t$. The vector $\vec k = \ve{k_1, k_2, \ldots, k_n}$ is nothing else but a list of non-negative  integers of length $n$. It determines which dots $\ovar$ are bound by the $\la$ in the following way: Consider all $\ovar$ in the term $t$ which are not bound in $t$ itself. Now, the first $k_1$ of these are not bound by the $\la$, the next one is, the following $k_2$ are again not bound and so on (see examples below). 

We denote the number of unbound $\ovar$ in an ordered preterm $t$ by $\freevars t$. Consequently, $\freevars \cdot$ is simply defined by: 
\[
\begin{array}{cccc}
\freevars x = 0, & \freevars \ovar = 1, & \freevars {\ApO t m u} = \freevars {t} + \freevars{u}, & \freevars {\LaO {\ve {k_1, \ldots, k_n}} t} = \freevars t - n.
\end{array}
\]
We call an ordered preterm $u$ an \define{ordered term} iff each sub-preterm $w$ of $u$ fulfils the following condition: If $w$ is an application $\ApO t m u$, the equation $m = \freevars t$ holds and if it is an abstraction $\LaO {\ve {k_1, \ldots, k_n}} t$, then $n + \sum_i k_i \leq \freevars t$ is satisfied. The latter condition states that if a $\la$ in $u$ binds a variable, this variable must actually exist while the first one just gives a meaning to the integer carried by an application. Clearly, any sub-preterm of an ordered term is again an ordered term. 
%
$u$ is called \define{closed} if $\freevars t = 0$.

Here are some examples of closed terms.
The $S$ combinator 
\[
\LaA x {\LaA y {\LaA z {\ApA x z \sspace \pa {\ApA y z}}}}
\]
would be written as 
\[\LaO {\ve 0} {\LaO {\ve 1} {\LaO {\ve{1,1}} {\ApO {\ApO \ovar 1 \ovar} 2 {\pa{\ApO \ovar 1 \ovar}}}}}
\]
Moreover, the term 
\[
(\LaA x {\LaA y {\ApA{\ApA a b} y}}) \sspace g \sspace f
\]
from Section~\ref{sec:example} would be represented as
\[
\ApO{\left(\ApO {\LaO{\emptyVec}{\LaO{\ve 0}{\ApO{\ApO{a}{0}{b}}{0}{\ovar}}}\right)}  0 f } 0 g
\]
(note that applications are still left-associative). We can see that the first $\la$ does not bind anything as it is annotated with the empty vector $\emptyVec$, while this is less obvious when it is written as $\la x$. 

At this point, we hope to have clarified the intended meaning of our syntax. A formal definition will be given in Section~\ref{sec:parsing}.

\section{Values and Evaluation} 
\label{sec:values}

Before specifying values and evaluation formally, we want to give an example to demonstrate how the information carried by a lambda should be used and why we always have exactly the needed information. Suppose we have the term
\[
\left(\LaO {\ve 0} {\LaO {\ve 1} {\LaO {\ve{1,1}} {\ovar \oapp 1 \ovar \oapp 2 {\pa{\ovar \oapp 1 \ovar}}}}}\right) \sspace g \sspace f \sspace n 
\]
that is, the $S$ combinator applied to three free variables
(we suppress the application indices $0$ for better readability). We want to get rid of the beta redexes, so we start by eliminating the first one. The outermost $\la$ is decorated with the vector $\ve 0$ of length one. Now, the single variable bound by this $\la$ should be replaced by $g$, so we start a substitution list and insert a single $g$:
\[
\left(\LaO {\ve 1} {\LaO {\ve{1,1}} {\ovar \oapp 1 \ovar \oapp 2 {\pa{\ovar \oapp 1 \ovar}}}}\right) \left[ g \right] \lspace f \sspace n
\]
The first remaining $\la$ is $\la^{\ve 1}$, so it does not bind the first variable (thus $g$ remains first in the substitution list), but the second one. Consequently, we add an $f$ after the $g$:
\[
\left(\LaO {\ve{1,1}} {\ovar \oapp 1 \ovar \oapp 2 {\pa{\ovar \oapp 1 \ovar}}}\right)  \left[ g, f \right] \lspace n
\]
Now the situation becomes more interesting. The only remaining $\la$ is now decorated with the vector $\ve{1,1}$, so one $n$ has to be inserted after the first entry ($g$) and another one must be placed after the subsequent entry ($f$):
\[
\left(\ovar \oapp 1 \ovar \oapp 2 {\pa{\ovar \oapp 1 \ovar}}\right)  \left[ g, n, f, n \right]
\]
All $\la$ have now been eliminated. The applications' indices tell us how the substitution list should be divided between the terms:
\[
\left(\ovar \oapp 1 \ovar\right)\left[ g, n \right] \lspace \left(\ovar \oapp 1 \ovar\right) \left[ f, n \right]
\]
We do the same step once more:
\[
\ovar[g]  \lspace \ovar[n]  \lspace \left(\ovar[f]  \lspace \ovar[n]\right)
\]
The only thing left to be done is to apply the substitutions in the obvious way:
\[
g \sspace n  \sspace (f \sspace n)
\]
Here, evaluation naturally leads to a term in beta normal form. This is not always the case: as an example, if we had tried to evaluate the above term without the $n$ (i.e. $S \oapp 0 f \oapp 0 g$), we would have got stuck at $\left(\LaO {\ve{1,1}} {\ovar \oapp 1 \ovar \oapp 2 {\pa{\ovar \oapp 1 \ovar}}}\right)  \left[ g, f \right]$. However, this would have been satisfactory as it would have shown that the term's normal form is an abstraction. In other words, our evaluation results in weak head normal forms.

Consequently, we define \define{values} in the following way:
\[ 
\begin{array}{lllrll@{\qquad}}
\mathsf{values}       & \ni & v,w & ::= & x \sspace \vec v & \mbox{large application} \\ 
			                   &&& \mid & \Val {\vec k} t {\vec v} & \mbox{closure} \\ 
\end{array}
\]
The \emph{large application}\footnote{
The argument vector $\vec v$ of a large application is sometimes
called a \emph{spine} \cite{cervesatoPfenning:spineCalculus}.  Large
applications $x\,\vec v$ also appear in the formulation of B\"ohm
trees \cite{barendregt:lambdacalculus}.}
 consists of a variable $x$ which is applied to a vector $\ve{v_1, v_2, \ldots, v_m}$ of values. It is to be read as a left-associative application, i.e. as $\pa{\pa{x \, v_1} v_2 \ldots } v_m$. 
Note that it is not necessarily ``large''. Quite the contrary, it
often only consists of the head (and the vector of values is empty).

A \emph{closure} $\Val {\vec k} t {\vec v}$ is the result of the evaluation process if the corresponding beta normal form of the term does not start with a free variable. The main part, $\LaO {\vec k} t$, is nothing other than a lambda abstraction in the syntax of ordered terms.
In addition, we need the substitution list $\vec v$ (which is simply a list of values) that satisfies $\length {\vec v} = \freevars{\LaO {\vec k} t}$. 
The idea is that the $i^{th}$ unbound $\ovar$ is to be replaced by $v_i$.  These substitution lists have already been used in the example above.

At this point, we want to introduce a notation for inserting a single item multiple times into a list. More precisely, if $\vec v = [v_1, v_2, \ldots, v_m]$ is a list, 
$\vec k = [k_1, k_2, \ldots, k_n]$ is a vector (i.e. also a list) of  non-negative integers satisfying $\sum_{i=1}^n k_i \leq m$ and $w$ is a single item, we write $\multiinsert {\vec v} {\vec k} {w}$  for the list that is constructed by inserting $w$ at each of the positions $k_1, k_1 + k_2, \ldots, \sum_{i=1}^n k_i$ into $\vec v$, i.e. for the list $[v_1, v_2, \ldots, v_{k_1}, w, v_{k_1 + 1}, \ldots, v_{k_1 + k_2}, w, v_{k_1 + k_2 + 1}, \ldots, v_m]$ (of course, it is possible that $\multiinsert {\vec v} {\vec k} {w}$ starts or ends with $w$).

We are now able to define the evaluation function $\ev {\cdot} {\cdot}$ which takes an ordered term $t$ as well as an ordered substitution list $\vec v$ and returns a value. The tuple must always satisfy the condition $\freevars t = \length {\vec v}$. In other words, the list carries neither too little nor redundant information.
At the start of the evaluation, the ordered substitution list is empty. 
Additionally, we specify the application $\cdot \ap \cdot$ of two values, which also returns a value and does not need anything else.
Our evaluation procedure uses a call-by-value strategy:

\[ 
\begin{array}{lcl@{\qquad} r}
 \ev x {\emptyVec} & = & x & \lab 1 \\ \noalign{\medskip}
 \ev \ovar {\ve{v_1}} & = & v_1 & \lab 2\\  \noalign{\medskip}
 \ev {t \oapp k u} {\ve{v_1, \ldots, v_n}} & = & {\ev t {\ve{v_1, \ldots, v_k}}} \ap {\ev u {\ve{v_{k+1}, \ldots, v_n}}}  & \lab 3 \\  \noalign{\medskip}
 \ev {\LaO {\vec k} t}  {\vec v} & = & \Val {\vec k} t {\vec v} & \lab 4 \\  
\\ 
 \pa{x \sspace \vec v} \ap w & = & x \sspace \ve{\vec v, w}  & \lab 5
\\ \noalign{\medskip}
 \Val{\vec k} t {\vec v}   \ap w & = & \ev t {\multiinsert{\vec v}{\vec k}{w}} & \lab 6
\end{array}
\]
First, if we want to evaluate a free variable \lab 1, the substitution list must be empty because of the invariant mentioned above. 
Second, in the case of a $\ovar$ \lab 2, the ordered list must have exactly one entry. This entry is the result of the evaluation. 
If we evaluate an application \lab 3, we evaluate the left and the right term. The application's index enables us to split the substitution list at the right position. Then, we have to apply the first result to the second. 
Evaluating an abstraction \lab 4 is easy. We just need to keep the substitutions to build a closure. 

If we want to apply a large application to a value $w$ \lab 5 , we just append $w$ to the vector of values (we write $\ve{\vec v, w}$ for $\ve{v_1, \ldots, v_n, w}$). 
The case of a closure $\Val {\vec k} t {\vec v}$ \lab 6 is less simple, but it is still quite clear what to do: $\vec k$ determines at which positions $w$ should be inserted in the ordered substitution list, so we just construct the list $\multiinsert {\vec v} {\vec k} w$. Then, $t$ is evaluated.

Concerning substitution lists, we talk about ``lists of values'' for
simplicity. More specifically, we want them to be lists of pointers to
avoid the duplication of ``real'' values during constructing lists
like $\multiinsert {\vec v} {\vec k} {w}$. 
Instead of simple linked lists,
we have also implemented these lists as dynamic functional arrays
represented as binary trees.  This reduces the asymptotic costs of
list splitting---$[v_1,\dots,v_n]$ to
$([v_1,\dots,v_k],[v_{k+1},\dots,v_n])$---
and multi-insertion $\multiinsert {\vec v} {\vec k} w$ 
from linear to logarithmic time (in terms of the length of the list
$\vec v$).  For an
experimental comparison of the two implementations
see Section~\ref{sec:experiments}.

\section{Parsing and Printing}
\label{sec:parsing}

In this section, we define how terms in normal syntax are translated into our ordered syntax (Parsing) and vice versa (Printing). 
To specify this, we need some notation. 
First of all, we write $\x$ for the set of variable names we want to use and $\te$ for the set of lambda terms in basic standard syntax (i.e. $\x \subset \te$, furthermore, $x \in \x$ together with $t, u \in \te$ implies $\ApA t u \in \te$ and $\LaA x t \in \te$). Additionally, $\ot$ is the set of terms in our ordered syntax defined above, $\otc$ the subset of closed ordered terms ($\freevars t = 0$) and $\va$ the set of values (defined in the previous section).
Moreover, we write $\xl$ for the set of lists of variable names and $\vl$ for the set of lists of values. 
For each set $\Gamma$ of variable names ($\Gamma \subseteq \x$), we denote the set of lists of elements of $\Gamma$ by $\gl$ and the ordered terms that do not contain any variable of $\Gamma$ (as a free variable) by $\otw$.

By writing $\ot \otimes \xl$ (resp. $\ot \otimes \vl$, $\otw \otimes \gl$, $\ldots$), we mean the subset $\{(t, \vec x) \ | \ \freevars t = \length {\vec x} \}$ of $\ot \times \xl$ (and analogous for the other cases).

For a finite 
set $\Gamma$ of variable names, we define the \define{correspondence relation}
${\parseb \cdot \Gamma \cdot \cdot ~ \subset ~ \te \times \ot \times
  \xl}$
(pronounce: ``corresponds in context $\Gamma$ to'').
The intuition is that ${\parseb M \Gamma u {\vec x}}$ means: 
$M$ is a term that corresponds to the ordered term $u$, where unbound
$\ovar$ are replaced by the (\emph{not} necessarily pairwise distinct)
variables in the list $\vec x$. 
The set $\Gamma$ can be seen as a filter that tells us which free variables do not occur in $u$ but in $\vec x$ instead.
\[
\begin{array}{c@{\qquad\qquad} c}
\infnamed{\quad (A)}{x \in \Gamma}{\parseb x \Gamma \ovar x}
&
\infnamed{\quad (B)}{ {\parseb M \Gamma t {\vec x}} \qquad 
        {\parseb N \Gamma u {\vec y}} \qquad {\length{\vec x} = m} } 
      {\parseb {M\,N} {\Gamma} {(\ApO t m u)}  {\vec x, \vec y} } 
\\ \\
\infnamed{\quad (C)}{x \not\in \Gamma}{\parseb x \Gamma x {}}
&
\infnamed{\quad (D)}{ 
         {\parseb M { (\Gamma \cup \{z\}) }  t {\multiinsert{\vec x}{\vec k}{z}   }} 
          \qquad 
          z \not\in \vec x  }
          {\parseb {\LaA z M} {\Gamma} {(\LaO {\vec{k}} t)} {\vec x}}
\end{array}
\]
It is important to note that $\parseb M \Gamma u {\vec x}$ implies that each variable occurring in $\vec x$ is contained in $\Gamma$ and each free variable occurring in $u$ is not contained in $\Gamma$. This can be shown by induction on $M$ (simultaneously for all sets $\Gamma$). 

By the same argument, one can see that (for each $M$ and $\Gamma$) there exists a unique tuple $(u, \vec x)$ satisfying $\parseb M \Gamma u  {\vec x}$, so we can consider $\parseFun \Gamma$ a function $\te \rightarrow \otw \times \gl$. Furthermore, we note that $\parseb M \Gamma u {\vec x}$ always implies $\freevars u = \length {\vec x}$. 
 
This also works the other way round. For each $\Gamma$ and each tuple
$(u, \vec x) \in \otw \otimes \gl$, there is (by induction on $u$) a
term $M \in \te$ satisfying $\parseb M \Gamma u {\vec x}$. Moreover,
this term $M$ is unique up to $\alpha$ equivalence. So, $\parseFun
\Gamma$ is actually a bijection between $\te / \alpha$ (the set of
$\alpha$ equivalence classes of terms) and $\otw \otimes \gl$. The
inference rules above show how to apply this bijection or its inverse
to a term or a tuple (in the last rule, any variable satisfying the
condition can be chosen for $z$), so we have 
a computable bijection $\te / \alpha \leftrightarrow \otw \otimes \gl$
determined by the rules for $\parseFun \Gamma$.

Choosing $\Gamma = \emptySet$, we get a bijection $\te/\alpha
\leftrightarrow \ot \otimes {\vec \emptySet}$.  As $\vec \emptySet$ is
only inhabited by the empty vector $\emptyVec$, we naturally get the
\define{parse} function $\parseFunE$ which maps $\te / \alpha$
bijectively on $\otc$.


The above construction also gives us a function $\otc \rightarrow
\te$, but this is not enough. We want to transform closures
(elements of $\ot \otimes \vl$) and values (elements of $\va$) 
into basic terms $\te$.
Therefore, we
define the two \define{print} functions $\prE : \ot \otimes \vl \rightarrow \te / \alpha$ and $\prValE : \va \rightarrow \te / \alpha$ simultaneously by recursion on the structure:
\[
\begin{array}{lclc}
\medskip
\pr{x}{\emptyVec} & = & x & \lab I \\
\medskip
\pr{\ovar}{[v]} & = & \prVal v & \lab {II} \\
\pr{t \oapp m u}{\vec v} & = & \ApA {\pr {t}{\vstart}  }  {\quad \pr
  {u}{\vrest}}  & \lab {III} \\
\medskip
&& \mbox{(split $\vec v$ at position $m$ to get $\vstart$ and $\vrest$)} \\
\pr{\LaO {\vec k} t}{\vec v} & = & \LaA z {\pr {t} {  \multiinsert{\vec v}{\vec k}{z}   }}   & \lab {IV}\\
\medskip
&& \mbox{where $z$ is any variable that does not occur freely in $t$ or $\vec v$ } \\
\prVal{x \sspace v_1 \sspace v_2 \sspace \ldots \sspace v_n} & = & x \ \prVal{v_1} \ \prVal{v_2} \ \ldots \ \prVal{v_n} & \lab V\\
\medskip && \mbox{(a large application simply becomes an application of terms)} \\
\prVal{\Val {\vec k} t {\vec v}} & = & \pr{\LaO {\vec k} t}{\vec v}  & \lab {VI}\\
\end{array}
\]
First, note that the printing functions are well-defined (i.e., they always terminate). This is because during evaluation of $\pr{u}{\vec v}$, we may safely assume that $\pr{t}{\vec w}$ is well-defined as long as $t$ is a strict subterm of $u$ and each value $w'$ in $\vec w$ is either only a variable (so termination of $\prVal {w'}$ is clear) or also in $\vec v$. Similar, during evaluation of $\prVal {x \sspace v_1 \sspace \ldots v_n}$, we may assume that $\prVal {v_i}$ is defined for each $i$.

For all $t \in \otc, M \in \te$, we have $\pr t \emptyVec = M$ if and only if $\parseb M \emptySet t { }$ (which just means $\parseFunE t = M$) as both judgements are defined identically in the case of closed ordered terms. This essentially (with implicit use of the bijection $\otc \leftrightarrow \ot \otimes \vec \emptySet$) means $\prE \circ \parseFunE  = id_{\te}$, i.e. the composition of parsing and printing is the identity.

\section{Correctness and Termination properties}
\label{sec:sound}

We still have not shown that our evaluation algorithm given in Section~\ref{sec:values} does not change the meaning of terms. The combination of parsing, evaluating and printing should never result in a term that is not beta equivalent to the original term. We also want to show a limited termination property.
To keep our argument simple, we just sketch the proofs and hope that the ideas become clear.

First, we attend to the correctness question. 
We need to convince ourselves that \emph{rewriting} according to the rules of the functions $\ev \cdot \cdot$ and $\ap$ does not cause an error. 
By \emph{rewriting}, we mean one step of the \emph{normal} or \emph{leftmost outermost} evaluation. We have demonstrated this in the example at the beginning of Section~\ref{sec:values}.
Printing should result in a term that is $\beta$ equivalent to the term we get if we rewrite before printing. 
This basically means that, for each evaluation rule on the left hand
side of the following table, we have to check that the equality on the
right hand side holds:

\begin{center}
\begin{tabular}{|   rcl  |  rcl    | c  }
\cline{1-6}
&&&&&& \\
$\ev x {\emptyVec} $&$=$&$ x    $ & $ \pr{x}{\emptyVec} $&$\, =_\beta \, $&$\prVal{x}$ & \qquad \lab 1 \\
&&&&&& \\
$\ev \ovar {\ve{v}} $&$=$&$ v   $ & $ \pr{\ovar}{v} $&$\, =_\beta \,$&$ \prVal v$  & \qquad \lab 2 \\
&&&&&& \\
$\ev {t \oapp k u} {[\vec v , \vec w]} $&$=$&$ {\ev t {\vec v}} \ap {\ev u {\vec w}}   $ & $ \pr {t \oapp k u} {[\vec v , \vec w]} $&$\, =_\beta \,$&$  \ApA {\pr{t}{\vec v}} {\pr{u}{\vec w}} $  & \qquad \lab 3 \\
&&&&&& \\
$\ev {\LaO {\vec k} t}  {\vec v} $&$=$&$ \Val {\vec k} t {\vec v} $ & $  \pr{\LaO {\vec k} t}{\vec v}$&$ \, =_\beta \,$&$  \prVal {\Val {\vec k} t {\vec v}} $  & \qquad \lab 4 \\  
&&&&&& \\
$\pa{x \sspace \vec v} \ap w $&$=$&$ x \sspace \ve{\vec v, w}   $ & $\ApA {\prVal {x \sspace \vec v} } {\prVal w}   $&$\, =_\beta \, $&$ \prVal{x \sspace \ve{\vec v, w}}$  & \qquad \lab 5\\
&&&&&& \\
$\Val{\vec k} t {\vec v}   \ap w $&$=$&$ \ev t {\multiinsert{\vec v}{\vec k}{w}}   $ & $ \ApA{\prVal{\Val{\vec k} t {\vec v}}}{\prVal w}  $&$ \, =_\beta \, $&$  \pr{t}{\multiinsert{\vec v}{\vec k}{w}}  $  & \qquad \lab 6\\
&&&&&& \\
\cline{1-6}
\end{tabular}
\end{center}

Note that ``rewriting using the evaluation rules'' results in expressions which are, more or less, a mixture of elements of $\ot \otimes \vl$ and $\va$. To be precise, such an expression is either in $\ot \otimes \vl$ or in $\va$ or a tuple (to read as simple application) of two expressions. The $1^{st}$, $2^{nd}$ and $4^{th}$ rule turn an element of $\ot \otimes \va$ into a value, the $3^{rd}$ turns it in a tuple of two such elements, and the last two rules turn a tuple of two values into one value or element. 
Although we do not define it formally, it should be clear how those expressions can be printed by using the printing functions for ordered terms and values (if an expression is a tuple, just print the function part and the argument part separately). In fact, while the function $\ev \cdot \cdot$ is formulated as a big step evaluation, the rewriting process can be understood as the corresponding small step (or one step) evaluation.

In the first five rows of the table, we only have to look at the definitions of the printing functions to see that the printed terms are not only $\beta$ equivalent but also equal. The very last rule $\lab 6$ requires closer examination: 
By rule $\lab {IV}$, the term $\prVal{\Val{\vec k} t {\vec v}}$ is equal to $\LaA z {\pr {t} {  \multiinsert{\vec v}{\vec k}{z}   }} $ for a (sufficiently) fresh variable $z$. Now, the definitions of the printing functions are ``context free'' in a way that guarantees  that $z$ occurs exactly $\length{\vec k}$ times (free) in $\pr {t} {  \multiinsert{\vec v}{\vec k}{z}   }$. Furthermore, replacing those occurrences by $\prVal{w}$ results in the term $\pr{t}{\multiinsert{\vec v}{\vec k}{w}}$. This means that, starting with 
$ \ApA{\prVal{\Val{\vec k} t {\vec v}}}{\prVal w}  $,
we have to use exactly one $\beta$ reduction step to get the term $  \pr{t}{\multiinsert{\vec v}{\vec k}{w}}  $.

As we have already seen that the composition of parsing a term and printing it afterwards does not change anything (up to $\alpha$ equivalence), we can now conclude that parsing, evaluating (a finite number of rewriting steps) and printing is equivalent to a number of $\beta$ reduction steps.

Now we discuss termination. Obviously, our evaluation function $\ev \cdot \cdot$ does not always terminate as some terms do not have a weak head normal form. However, $\ev \cdot \cdot$ terminates whenever it is applied to ($\parseFunE t$) if the usual $\beta$ reduction is strongly normalizing on $t$. 
The main consequence of this is that evaluation terminates for all well-typed terms. 
To prove this statement, assume that there is such a term $s_0 \in \te$ so that the evaluation of $t_0 := \parseFunE s_0$ does not terminate. 
Then, we get an infinite sequence $t_0, t_1, t_2, \ldots$ where $t_{i+1}$ is the result of rewriting (a subexpression of) $t_i$ using one of the evaluation rules. If we print $t_0, t_1, t_2, \ldots$, we get a sequence $s_0, s_1, s_2, \ldots$ of terms in $\te$, where $s_{i+1}$ is either ($\alpha$) equal to $s_i$ or arises from $s_i$ in exactly one $\beta$ reduction step. If $\beta$ reduction is strongly normalizing on $s_0$, the sequence has to become constant at some point, i.e. $s_N = s_{N+1} = s_{N+2} = \ldots$ for some $N$. This implies that, after the first $N$ rewriting steps, rule \lab 6 is not used anymore. Define the \emph{weight} $w(t)$ of an expression $t$ to be $1$, if the expression is just an element of $\ot \otimes \va$, to be $2$, if it is a value of the \emph{closure} type, to be 
$1 + 2^n + w(v_1) + w(v_2) + \ldots + w(v_n)$, if it is a value of the form $x \, v_1 v_2 \ldots v_n$ (i.e. a \emph{large application}) and, if it is a tuple of two expressions, as the sum of both weights. Then, each of the rewritings that are induced by the first five lines in the table increase the weight of the expression, so we get $w(t_N) < w(t_{N+1}) < w(t_{N+2}) < \ldots$;  however, as the total number of values (and tuples in $\ot \otimes \va$) is bound by the length of the term we get after printing $t_N$ (or any $t_{N+i}$), the sequence is bounded, resulting in the required contradiction.

\section{Experiments and Results}
\label{sec:experiments}

The specified term representation and evaluation have been implemented
in Haskell. They have been used by a type checker to check large files
of dependently
typed terms of the Edinburgh Logical Framework which were kindly
provided by Andrew W. Appel (Princeton University). To make this
possible, an extended syntax has been used that includes $\Pi$-types
, constants and definitions. It is straightforward to expand our
evaluation algorithm to the extended syntax---for details consult the
Bachelor's thesis of the second author \cite{kraus:bachelor}. 
The substitution lists
have been implemented as simple Haskell lists, and also as balanced
binary trees (following Adams \cite{adams:jfp93})
for better asymptotic complexity.  Both variants were
evaluated for performance 
[referred to as \emph{Ordered (trees)} and \emph{Ordered (lists)}].

For comparison, the completely analogous algorithm for terms in
extended basic syntax (i.e. $\te$) has been used [\emph{Simple
  Closures}]. Furthermore, we have tested a strategy that always
evaluates completely (i.e. produces $\beta$ normal forms) using
Hereditary Substitution [\emph{Beta Normal Values}]
\cite{watkins:concurrentlftr}.

Our main test file \textsf{w32\_sig\_semant.elf} with a size of
approximately 21 megabytes contains a proof described in
\cite{appel:toplas10}. We also tested smaller parts of this file, more
precisely, the first $6000$, $10,000$ and $12,000$ lines without the
rest (named \textsf{6000.elf} and so on). Later terms tend to be
larger, so the tests with fewer lines needed much less time.

All tests were executed on the same server 
\mbox{\texttt{baerentatze.cip.ifi.lmu.de}}
working with a CPU of type \emph{AMD Phenom II X4 B95} (only
one core used, 3 GHz) and 8 GB system memory. The measurements
of space and time consumption are given in the following tables (rounded
average values). More specifically, \emph{time} refers to the total time, including parsing the input file and transforming (if necessary) the representation to our ordered representation or to one that uses De Bruijn terms. However, in our tests, the transformation process only took a negligible amount of time. The time value does, however, not include any printing of the terms or the values (with printing, the total time increased significantly). \emph{Space} refers to the peak space usage of the whole process.

\begin{center}
\begin{tabular}{| l || c | c |}
\multicolumn{3}{c}{\textsf{6000.elf} (file size: 3.8 MB)}\\
\hline
& time (sec) & space (MB) \\
\hline
\hline
Ordered (trees) & 18.9 & 1111 \\
\hline
Ordered (lists) & 18.6 & 1114\\
\hline
Simple Closures & 18.5 & 1152\\
\hline
Beta Normal Values & 27.6 & 2034\\
\hline
\end{tabular}

\vspace{5pt}

\begin{tabular}{| l || c | c |}
\multicolumn{3}{c}{\textsf{10000.elf} (file size: 12.9 MB)}\\
\hline
& time (sec) & space (MB) \\
\hline
\hline
Ordered (trees) & 61.0 & 3230\\
\hline
Ordered (lists) & 60.6 & 3237\\
\hline
Simple Closures & 60.0 & 3302\\
\hline
Beta Normal Values & 98.7 & 5878\\
\hline
\end{tabular}

\vspace{5pt}

\begin{tabular}{| l || c | c |}
\multicolumn{3}{c}{\textsf{12000.elf} (file size: 17.8 MB)}\\
\hline
& time (sec) & space (MB) \\
\hline
\hline
Ordered (trees) & 84.3 & 5096 \\
\hline
Ordered (lists) & 83.8 & 5103 \\
\hline
Simple Closures & 83.6 & 5226 \\
\hline
Beta Normal Values & 137.7 & 8513 \\
\hline
\end{tabular}
\end{center}

Unsurprisingly, beta normal values perform significantly worse than each of the other possibilities. However, the difference is smaller than it could have been expected. This might be due to the fact that during type checking, total evaluation of a term is often necessary anyway, thereby reducing the hereditary substitution's disadvantage. 

Although none of the other strategies exhibited any shortcomings in the comparisons above, the following results for the complete file are remarkable. Here, implementing ordered substitutions as normal Haskell lists seems to be much more efficient than using tree structures: 

\begin{center}
 \begin{tabular}{| l || c | c |}
\multicolumn{3}{c}{\textsf{w32\_sig\_semant.elf} (file size: 20.9 MB)}\\
\hline
& time (sec) & space (MB) \\
\hline
\hline
Ordered (trees) & 108.4 & 8877\\
\hline
Ordered (lists) & 94.8 & 4948\\ 
\hline
Simple Closures & 94.3 & 5068 \\ 
\hline
Beta Normal Values & 169.8 & 9044  \\
\hline
\end{tabular}
\end{center}

Our Simple Closures are still on the same level as Ordered Representation with lists, but the trees are far behind. 
In comparison, the type checker of the Twelf project, \emph{Twelf
  r1697} (written in \emph{Standard ML} and compiled with
\emph{MLton}'s whole program optimizations \cite{fluetWeeks:icfp01}) 
does the job nearly five
times faster while using only 2720 megabytes of
memory. 


\section{Related Work and Conclusions}
\label{sec:concl}

Our term representation is inspired by intuitionistic implicational
linear logic in natural deduction style which has explicit operations
for weakening and contraction
\cite{bentonBiermanDePaivaHyland:tlca93}.  With explicit weakening and
contraction, one easily maintains complete information about the free
variables of a term at each node \cite{kesnerLengrand:infcomp07}.  Our
term representation incorporates weakening and contraction into lambda
abstraction.  By using inspiration from ordered logic, we reduce the
stored information at application nodes to a minimum, namely an
integer; further, our variable nodes need to carry no information at all.

Another means to maintain information about free variables are
\emph{director strings} by Sinot \cite{sinot:jlc05}.  Application
nodes come with a map that tell for each variable whether it appears
in the left or the right subterm or in both.  Our term representation
can be seen as an optimized version of director strings, however, we
have no experimental comparison. 
Sinot \etal~\cite{fernandezMackieSinot:aaecc05} present some
performance results of director strings; however, it is restricted
to evaluation of some specific big lambda-terms.  There is no study on
their relative performance in a realistic application---yet that is
our concern.

An alternative to explicit substitutions is Nadathur's suspension
calculus \cite{nadathur:jflp99}, which, in a refinement, also
maintains information about closedness of subterms.  In this
refinement, the suspension calculus maintains at least partial
information about the free variables of a subterm.  
As the basis of an implementation of
$\lambda$-Prolog \cite{nadathur:flops01}, 
Nadathur has proven the efficiency of his term
representation not only for normalization and equality checking, but
also for higher-order unification.

Building on the suspension calculus,
Liang, Nadathur, and Qi \cite{liangNadathurQi:jar05} have evaluated
different term representations in the context of $\lambda$Prolog, a
study that compares to our study of term representations for the
Edinburgh Logical Framework.  They have tested different combinations
of features, confirming our result that lazy substitution is 
preferable to eager substitution {[Beta Normal Forms]},
even more so when several substitutions are gathered into one
traversal {[Closures, Ordered]}.  They also test a variant where
each term is equipped with an \emph{annotation}, a flag telling whether
this term is open, \ie, has free variables, or closed, \ie, has no
free variables.  In their experimental evaluation, these annotations
pay off greatly for the poorly behaving eager substitution, yet give
negligible advantage for explicit substitutions.  It is
hypothesized that in a combined substitution, each subterm will
mention at least one variable with a high probability, so the
traversal has to run over most of the whole term---this is certainly
different in the substitution for a single variable.

To summarize, we have presented a new term representation for the
lambda-calculus inspired by ordered linear logic, and experimentally
compared it with well-known representations (closures, normal forms)
in a prototypical implementation of a type checker for the Edinburgh
Logical Framework.  The experiments were carried out on large realistic
proof terms, constructed manually and mechanically.  

The results were not significantly in favor of our new representation.
This might be due to the application domain, LF signature checking.
For one, LF-definitions are closed, which means that substitutions
never need to traverse a definition body when the definition is
expanded, and this optimization is shared by all the term
representations we compared.  Secondly, we only tested type checking,
not type reconstruction via unification.  During type checking, where
equality tests are expected to succeed, full normal forms are always
computed, and closures are very short-lived in memory.  More space
leaks are to be expected in applications such as logic programming or
type reconstruction, where unification is needed, which is not
expected to always succeed.   In constraint-based unification,
unsolvable constraints might be postponed, keeping closures alive for
longer.  In such situations, the benefits of our representation might
be more noticeable, more experiments are required.

In the future, we plan to investigate further term representations
such as term graphs, and perform more experiments.  The literature on
experimentally successful term representations is sparse, our work
contributes to close this gap.  Our long term goal is to find a
term representation which speeds up Agda's type reconstruction.

\paragraph*{Acknowledgements}
The idea for the here presented ordered term representation was
planted in a discussion with Christophe Raffalli who invited the first
author to Chambery in February 2010.  He mentioned to me that in his
library \texttt{bindlib} formation of closures restricts the
environment to the variables actually occurring free in the code.  I
also benefited from discussions with Brigitte Pientka and Stefan Monnier.

Thanks to Gabriel Scherer for comments on a draft version of this
paper and pointers to related work. 
The final version of this paper was stylistically improved with the highly appreciated help of Neil Sculthorpe.

\bibliographystyle{eptcs}
\bibliography{nonauto-lfmtp11}

\end{document}